\documentclass[osajnl,twocolumn,showpacs,superscriptaddress,10pt]{revtex4-1}
\usepackage{amsmath,amssymb,graphicx}
\begin{document}

\title{Broadband quasi-phase matching in MgO:PPLN thin film}

\author{Licheng Ge}
\author{Yuping Chen}\email{Corresponding author: ypchen@sjtu.edu.cn}
\author{Haowei Jiang}
\author{Guangzhen Li}
\author{Bing Zhu}
\author{Yi'an liu}
\author{Xianfeng Chen}\email{xfchen@sjtu.edu.cn}
\affiliation{State Key Laboratory of Advanced Optical Communication Systems and Networks, School of Physics and Astronomy, Shanghai Jiao Tong University, 800 Dongchuan Road, Shanghai 200240, China}

\begin{abstract}
Future quantum information networks operated on telecom channels require qubit transfer between different wavelengths while preserving quantum coherence and entanglement. Qubit transfer is a nonlinear optical process, but currently the types of atoms used for quantum information processing and storage are limited by the narrow bandwidth of up-conversion available. Here we present the first experimental demonstration of broadband and high-efficiency
quasi-phase matching second harmonic generation (SHG) in a chip-scale periodically poled lithium niobate thin film. We achieve large bandwidth of up to 2 THz for SHG by satisfying quasi-phase matching and group-velocity matching simultaneously. Furthermore, by changing film thickness, the central wavelength of quasi-phase matching SHG bandwidth can be modulated from 2.70 $\mu$m to 1.44 $\mu$m. The reconfigurable quasi-phase matching lithium niobate thin film provides a significant on-chip integrated platform for photonics and quantum optics.

\bigskip
\noindent\textbf{\textit{OCIS Codes:(130.3730)}} Lithium niobate; (310.6845) Thin film devices and applications; (190.2620) harmonic generation and mixing; (190.4390) Nonlinear optics, integrated optics.
\end{abstract}


\maketitle 

\begin{figure}[htbp]
\centering
\includegraphics[width=8 cm]{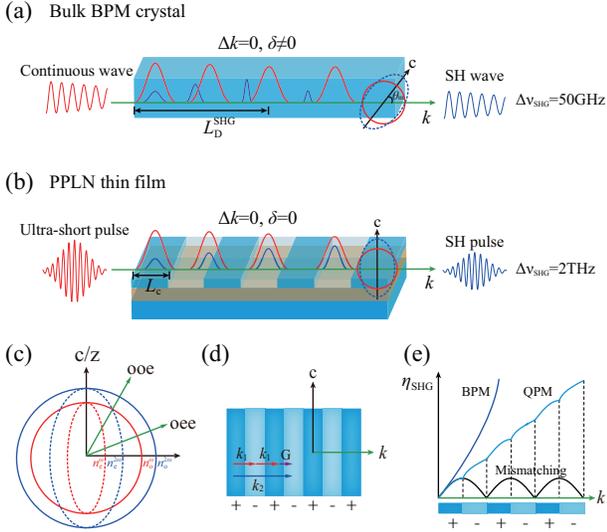}
\caption{(a) Traditional GVMM up-conversion ($\delta\neq0$) in bulk birefringent-phase-matching (BPM) crystal with 50 GHz bandwidth, only supporting continuous wave up-conversion. (b) GVM up-conversion ($\delta$=0) in PPLN thin film with 2 THz bandwidth, supporting ultra-short pulse up-conversion. $L_\text{c}$ is the coherence length. (c) Diagram of birefringent-phase-matching of negative uniaxial crystal. $\theta_\text{m}$ is the phase-matching angle between the optical axis c and the wave vector $k$. (d) Scheme of QPM by offering a reciprocal vector G. (e) The efficiencies of SHG for different phase-matching types.}
\label{fig1}
\end{figure}

\begin{figure}[htbp]
\centering
\includegraphics[width=8 cm]{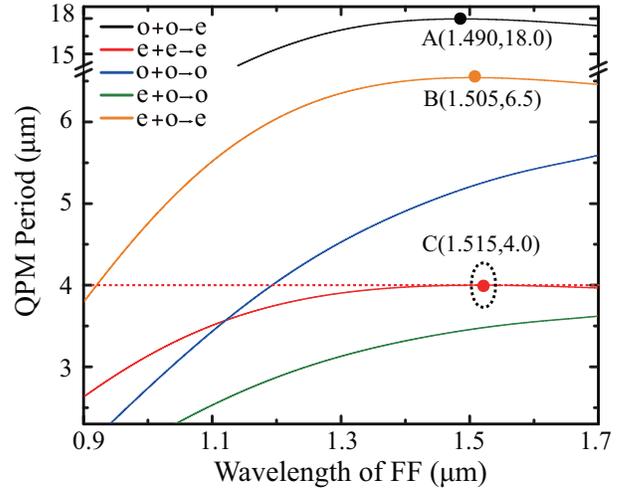}
\caption{ Simulated QPM period (2$L_\text{c}$ ) as a function of fundamental wavelength for different types of  up-conversions. GVM occurs at the extreme of the dispersion curve, in which case it exists in three kinds of up-conversion, as marked as points A(1.490,18.0), B(1.505,6.5) and C(1.515,4.0). }
\label{fig2}
\end{figure}

\begin{figure}[htbp]
\centering
\includegraphics[width=8 cm]{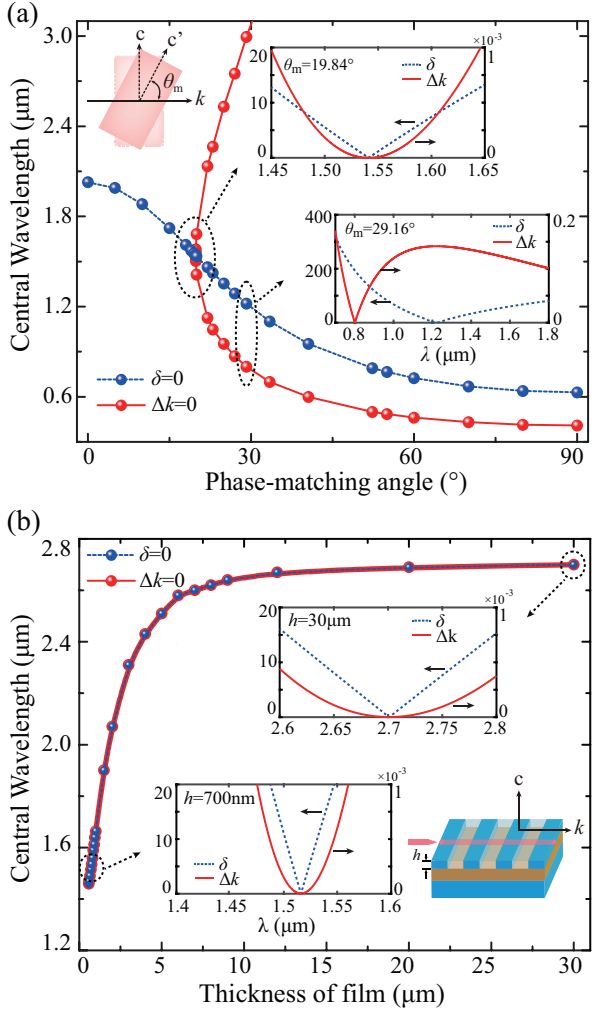}
\caption{(a) BBO: Central wavelengths of $\delta$=0 and $\Delta k$=0 as a function of $\theta_\text{m}$. $\delta$=0 and $\Delta k$=0 simultaneously only occurs at $\theta_\text{m}$=19.84$^{\circ}$, as inserted in (a). At other angles only one matching type can be satisfied. (b) PPLN thin film: Central wavelengths of $\delta$=0 and $\Delta k$=0 as a function of the thickness of film.  $\Delta k$=0  can always be satisfied under the same condition when $\delta$=0  if given a proper QPM period. Insert, two specific examples that $\delta$ and $\Delta k$ equal to zero the same time, when $h$=30 $\mu$m at $\lambda$=2.7 $\mu$m (nearly bulk) and $h$=700 nm at $\lambda$=1.515 $\mu$m, respectively.}
\label{fig3}
\end{figure}

Infrared entangled photons have played an essential role in quantum metrology, computing and imaging \cite{milburn1989quantum,knill2001scheme,duan2004scalable,lugiato2002quantum}. Nonlinear up-conversion is a common way to convert infrared light to the visible regime for detection \cite{brustlein2007absolute,dam2010high,nee2007two,huang2012few}, but there still remains some obstacles. The low conversion efficiency \cite{falk1978internal}  leads to weak nonlinear interactions between the entangled photons, which can be enhanced with ultrahigh photon flux by generating broadband entangled pairs \cite{dayan2005nonlinear}. Secondly, for time-resolved single-photon detection with femtosecond resolution \cite{kuzucu2008time}, it requires detectors to have a broad bandwidth. Thirdly, high brightness and widely tunable single-photon source is significant in quantum interference, which has been achieved by elimination of spectral entanglement though GVM \cite{almendros2009bandwidth,evans2010bright}. To meet the demand for increasing data rates, it is highly desired for integrated optics to combine all the photonic components including light sources, computing units, modulators and detectors on a single chip \cite{leng2011chip,jin2014chip}. So far, no report has been focused on the realization of tunable broadband SHG on chip-scale.
Recently, LiNbO$_3$ (LN) thin film has attracted much interest \cite{poberaj2012lithium} and has been commercially developed for electro-optical modulators \cite{huang2014helium,cai2016electric} and microdisk resonators \cite{wang2014integrated,lin2015fabrication}. Traditional waveguide, such as proton exchange waveguide, has small index contrast ($\Delta n$=0.09) and only supports TM mode \cite{sun2012466}. By using ion implantation and wafer bonding, LN thin film exhibits high nonlinear and electro-optic coefficients, low intrinsic absorption loss and large transparent windows as bulk materials have been reported \cite{fejer1992quasi}. The good mode confinement is due to the high refractive index contrast between LN and SiO$_2$ ($\Delta n$=0.75). Therefore, LN thin film can be a promising candidate for the highly integrated photonic platform with infrared sources and detectors on a single chip. However, previously reported LN thin film-based upconverters still have some limitation. One is the poor mode overlap in LN thin film waveguide \cite{geiss2015fabrication}, which can be solved by QPM technology \cite{armstrong1962interactions,chan2011synthesis,qin2008wave}. Besides, all the reported LN thin film-based upconverters have narrow bandwidth of 50 GHz due to group-velocity-mismatching (GVMM, $\delta\neq0$, see Fig.~\ref{fig1}(a)), limiting its applications on broadband up-conversion for single-wavelength wave and ultrafast pulses \cite{gong2010all,zhang2008flexible,dougherty1994dual,heilweil1989ultrashort}.
In this work, we demonstrated the first tunable broadband upconverter by achieving GVM ($\delta$=0) on chip-scale periodically poled lithium niobate (PPLN)  thin film. GVM can always be satisfied under the same condition with QPM if given a proper QPM period as introduced in Fig.~\ref{fig1}(b). As a result, the bandwidth can be as large as 2 THz. In addition, the dispersion condition for satisfying QPM and GVM simultaneously can be largely modulated from 2.70 $\mu$m to 1.44 $\mu$m via changing the PPLN film thickness. Our fabricated crystal film supports both TE and TM  modes \cite{poberaj2012lithium}  so that different types of up-conversion are allowed. The dimension of the crystal film can be a few centimeters long, which leads to the great enhancement of conversion efficiency. Experimentally, we observed TM$^{\omega}+$TM$^{\omega}\rightarrow$TM$^{2\omega}$  type broadband up-conversion process by using the largest nonlinear coefficient $d_{33}$ (27.2 pm/V). The sample is 4 cm long and has 700 nm-thick PPLN thin film with the fifth-order QPM period of 20 $\mu$m. The measured up-conversion bandwidth is 1.875 THz, and normalized conversion efficiency is 3.3 \%/W. The efficiency can be enhanced to 82.5 \%/W theoretically if using the first-order QPM period (4 $\mu$m).
Supposing $v_1$ and $v_2$ are the group velocities of fundamental-frequency (FF) and second-harmonic (SH) waves, the  spectral intensity of SH-pulse is expressed as \cite{Rulli1998Femtosecond}
\begin{equation}
  S_2(\omega,L)\propto \text{sinc}^2\{[(v_2^{-1}-v_1^{-1})\omega-\Delta k]L/2\}I_1^2.
\label{eq1}
\end{equation}
It shows that the two main factors that affect the spectral bandwidth are group-velocity-mismatching ($\delta=v_2^{-1}-v_1^{-1}$) and wave-vector-mismatching $\Delta k$. $L$ and $I_1$ are the length of the crystal and the intensity of FH wave, respectively. Considering $\Delta k$=0, the GVMM will cause the  temporal walk-off effect between the FH and SH pulses. The SHG process will be stopped if the two pulses completely separate in time. The walk-off distance is expressed as $L_D^{\text{SHG}}=\tau_{p1}/|v_2^{-1}-v_1^{-1}|$ (Fig.~\ref{fig1}(a)), where $\tau_{p1}$ is the pulse width of FF wave. GVMM can be ignored only when  $L\ll L_D^{\text{SHG}}$. The dependence of wave-vector-mismatching and GVMM is derived as \cite{yu2002broadband}
\begin{equation}
\frac{d(\Delta k)}{d\lambda}=\frac{4\pi c}{\lambda^2}\delta.
\label{eq2}
\end{equation}
It shows that GVM ($\delta$=0) can be achieved in the spectral region where the wave-vector mismatching takes an extremum [$d(\Delta k)/d\lambda/$=0]  around which the coherence length (or the QPM period) is nearly stand over a wide wavelength range.
For the QPM up-conversion, the wave-vector-mismatching is given by $\displaystyle\Delta k=(n^{2\omega}-n^{\omega}){4\pi/\lambda-2\pi}/\Lambda$ with the domain period of $\Lambda$ (=$2L_c$).  $n^{\omega}$ and $n^{2\omega}$ are the indices of FF and SH wave. Refractive indices are calculated by Sellmeier equations \cite{gayer2008temperature}.
In the PPLN thin film, the index should be replaced by  effective refractive index obtained by  solving the dispersion relationship of planar waveguide \cite{snyder2012optical}. Then the material dispersion  is plotted as the relationship between   QPM period ($\Lambda$) and fundamental wavelength (Fig.~\ref{fig2}). It has been reported in bulk QPM LN at the telecommunication band by using  type-I up-conversion (ooe) \cite{yu2002broadband,gong2010all,zhang2008flexible}. For the famous second-harmonic generation (SHG) crystal,  the  birefringent-phase-matching (BPM) condition of beta barium borate (BBO) can only be satisfied under a specific phase-matching angle $\theta_\text{m}$, therefore it only supports two kinds of SHG, including o+o$\rightarrow$e and  o+e$\rightarrow$e (see Fig.~\ref{fig1}(c)). The QPM  technique \cite{armstrong1962interactions} used in LiNbO$_3$ can avoid the difficulty of angle-phase-matching by  offering a  reciprocal vector G  as shown in Fig.~\ref{fig1} (d)-(e) and therefore support several kinds of up-conversion process.
Then we studied the conditions of obtaining simultaneous QPM and GVM in the thin film. For angle-phase-matching SHG in BBO crystal, phase-matching (PM) should be satisfied first, where the phase-matching angle $\theta_{\text{m}}$ changes monotonously with the FF wavelength. Figure \ref{fig3} shows the central wavelengths of PM and GVM as a function of $\theta_{\text{m}}$ in BBO crystal. GVM and PM can occur simultaneously at $\theta_{\text{m}}=19.84^{\circ}$. At other angles only one of them is satisfied. For example, for $\theta_\text{m}$=29.16$^{\circ}$,  $\Delta k$=0  but $\delta$=187 fs/mm, which means in order to avoid group-velocity-mismatching for 20 fs pulses, the length of the crystal should be less than 107 $\mu$m. While for PPLN thin film, GVM is automatically satisfied (see Fig.~\ref{fig3}(b)). The central wavelength can be tuned simply by changing the film thickness. The modulation range for central wavelength is as large as 1260 nm (1.44$\sim$2.70$\mu$m), corresponding to the film thickness change of 30 $\mu$m. The bandwidth of up-conversion in PPLN thin film is larger than 2 THz.
We fabricated the sample by starting with a periodically poled substrate instead of single-domain LiNbO$_3$
substrate. First, a z-cut   LiNbO$_3$ wafer of 3" diameter with 20 $\mu$m QPM period  is He-ions implanted to a required depth. Another silicon handle sample coated by 2 $\mu$m-thickness SiO$_2$ layer was surface polished to 0.35 nm roughness enabling direct wafer bonding. The bonded pair of samples is then annealed  to improve the bonding strength. By a further increase of the temperature, the sample splits along the He implanted layer. Afterwards, it is annealed again before the sample surface is polished to 0.5 nm  roughness by another chemical mechanical polishing process.
\begin{figure}[htbp]
\centering
\includegraphics[width=8 cm]{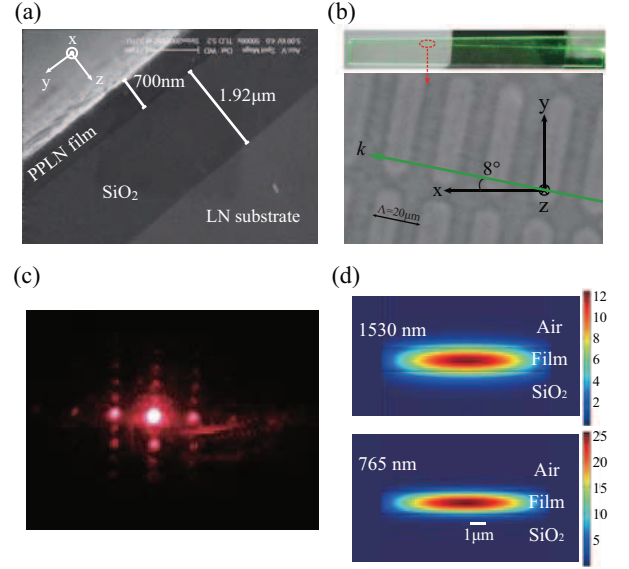}
\caption{ (a) SEM image of the endface of the  sample. A 700 nm-thick PPLN thin film is sitting on a SiO$_2$ layer with a LN substrate. (b) Observed light confinement and top-view of the periodic QPM structure.The direction of the periodic structure has a 8$^{\circ}$ angle-off with  x axis. (c) Diffraction pattern of PPLN thin film which indicates the grating structure on the interface of PPLN and SiO$_2$. (d) simulated intensity distributions of fundamental and second harmonic waves with TM mode. }
\label{fig4}
\end{figure}
Experimentally, we used a 700 nm-thick, 20 $\mu$m-QPM period and 4 cm-long PPLN thin film to full the GVM up-conversion process. Figure \ref{fig4}(a) and \ref{fig4}(b) show the side view and top view SEM images of the sample, respectively. Then a tunable continuous laser (1518-1618 nm) served as the pump source amplified by erbium-doped fiber amplifiers, where the polarizations were controlled by the polarization controller. The light then injected to the sample by a lensed fiber, which was set on a XYZ-translation stage with a resolution of 50 nm. The sample was set on YZ-translation stage with a temperature controller at an accuracy of 0.1 $^{\circ}$C. The output light was collected by a 20$\times$ objective lens and the generated SH wave was monitored by an optical spectrum analyzer (200-1100 nm). We launched a green laser tracer to the sample at the normal direction but observed the light propagation has about 8$^{\circ}$ misalignment with respect to x axis (see Fig.~\ref{fig4}(b)). This angle agrees with the QPM grating direction.
Since the QPM grating is periodically discontinues along y axis as well, it proves that light is confined laterally by this discontinuity \cite{kim2005optical,mhaouech2016low,jin2015mode}. We can see the diffraction pattern in Fig.~\ref{fig4}(c) when a red laser is shot perpendicularly to the surface, which indicates the grating structure on the interface of PPLN film and SiO$_2$. COMSOL simulated intensity distribution of FF and SH waves for TM mode is plotted in Fig.~\ref{fig4}(d). With QPM period of 20 $\mu$m, our sample only supports e+e$\rightarrow$e type GVM up-conversion (red line and point C in Fig.~\ref{fig2}) at the decreased fifth-order QPM conversion efficiency. The optimized temperature for GVM up-conversion at the wavelength of 1530 nm is 24.1 $^{\circ}$C. We verified the FF and SH wave propagating in the PPLN thin film are both in TM mode. The converted SH power is 120 pW at the pump power of 60 $\mu$W, leading to the normalized efficiency of 3.3 \%/W. It is worth noting that the efficiency can be enhanced to 82.5 \%/W  theoretically if using the first-order-QPM period (4 $\mu$m).
\begin{figure}[htbp]
\centering
\includegraphics[width=8.6 cm]{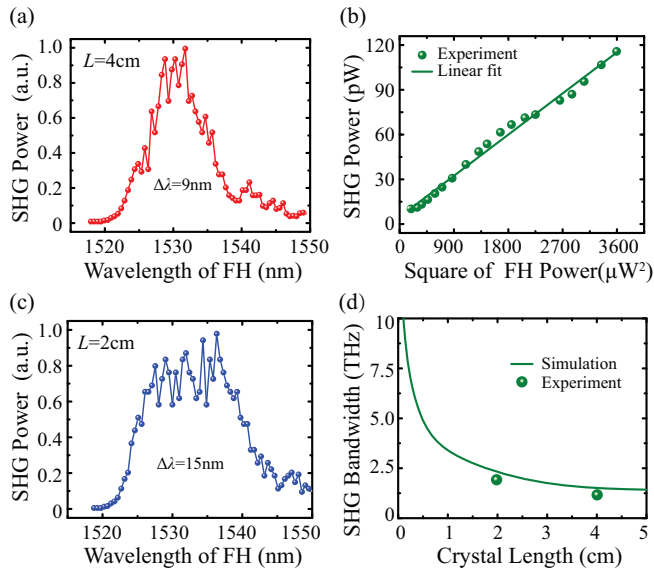}
\caption{ (a) Measured normalized SHG power  as a function of fundamental wavelength. The  upconversion bandwidth is 9 nm (1.125 THz) for a 4 cm-long crystal.  (b) Linear relationship between SHG power and the square of input FH power at wavelength of 1530 nm and temperature of 24.1 $^{\circ}$C.  (c) Recorded upconversion bandwidth for a 2 cm-long crystal, which is 15 nm (1.875 THz). (d) Experiment and simulation upconversion bandwidths as a function of crystal length.}
\label{fig5}
\end{figure}
The SH power as a function of FF wavelength is plotted in Fig.~\ref{fig5}(a), from which the up-conversion bandwidth is calculated to be 9 nm or 1.125 THz. Figure \ref{fig5}(b) demonstrates the linear relationship between the measured SH power and the square of FF power \cite{boyd2003nonlinear}. Reducing the interaction length can further broaden the up-conversion bandwidth but will result in lower conversion efficiency. In another experiment, a 2 cm-long sample was tested to provide 15 nm or 1.875 THz bandwidth(see Fig.~\ref{fig5}(c)). We simulated in Fig.~\ref{fig5}(d) the theoretical SHG bandwidth as a function of crystal length, and compare it with the experimental results. If using 1 cm-long crystal, the bandwidth can be as broad as 3.2 THz. We notice the shift of experimental central wavelength (1530 nm) from the theoretical value (1515 nm). This is due to the inaccuracy of Sellmeier equations \cite{gayer2008temperature} and the idealization of slab waveguide used in the simulation. The fifth order QPM used in our experiment results in low conversion efficiency, but it can be improved by fabricating first order QPM period of 4 $\mu$m.
In conclusion, we demonstrated the first tunable broadband up-conversion on thin periodically poled lithium niobate thin film by satisfying group-velocity-matching and quasi-phase-matching simultaneously. The central wavelength of the up-conversion band can be largely tuned from 2.7 $\mu$m to 1.44 $\mu$m by changing the film thickness. The measured bandwidth is 1.875 THz for a 2 cm-long sample, along with the conversion efficiency of 3.3 \%/W. The efficiency can be further enhanced by using first order QPM period and the bandwidth can be broader by shorter interaction length. Our demonstration could lead to the integration of infrared sources and detectors on a single chip, and other more advanced photonic integration devices that require broadband operation, for example, flexible wavelength converter, all optical wavelength broadcast, and $\chi^{(2)}:\chi^{(2)}$ cascading all-optical switch.

This work is supported by the National Natural Science Foundation of China (NSFC) (11574208),the National Key R $\&$ D Program of China (2017YFA0303700).


\end{document}